\documentstyle[epsfig,aps,prb,multicol]{revtex}

\draft

\begin{document}

\title{Effect of Dilution on First Order Transitions:\\
The Three Dimensional Three States Potts Model}
\author{H. G. Ballesteros$^1$, L. A. Fern\'andez$^1$,
V. Mart\'{\i}n-Mayor$^{1,2}$, \\
A. Mu\~noz Sudupe$^1$, G. Parisi$^2$ and J. J. Ruiz-Lorenzo$^1$.
\vspace{0.3 cm}}
\address{$^1$ Departamento de F\'{\i}sica Te\'orica I, 
Universidad Complutense de Madrid, 28040 Madrid. Spain.}
\address{$^2$ Dipartimento di Fisica, Universit\`a  di Roma I ``La Sapienza''
and INFN,
00185 Roma. Italy.\\
{\small \bf e-mail: \tt hector@lattice.fis.ucm.es,
laf@lattice.fis.ucm.es, victor@chimera.roma1.infn.it,
sudupe@lattice.fis.ucm.es, giorgio.parisi@roma1.infn.it, 
ruiz@lattice.fis.ucm.es.}
}

\date{June 14, 1999}

\maketitle

\begin{abstract}

We have studied numerically the effect of quenched site dilution on a
first order phase transition in three dimensions. We
have simulated the site diluted three states Potts model studying in
detail the second order region of its phase diagram.  We have found
that the $\nu$ exponent is compatible with the one of the three
dimensional diluted Ising model whereas the $\eta$ exponent is
definitely different.
\end{abstract}

\begin{multicols}{2}
\narrowtext

\section{Introduction}

The study of the impurities effect on the critical behavior of a pure
material is an important issue since frequently real systems cannot be
considered as pure. Nowadays the effect of dilution (disorder coupled
to the energy density) on second order phase transitions is well
understood. The Harris' criterion~\cite{Harris} states that if the
specific-heat of the pure system presents a power-like divergence (i.e. 
$\alpha_{\mathrm{pure}}>0$) the disorder induces a new
Universality Class. Otherwise ($\alpha_{\mathrm{pure}}<0$) the
critical behavior of the model remains unchanged. The criterium does
not decide in the marginal case $\alpha_{\mathrm{pure}}=0$.  Moreover,
it is possible to show rigorously that for all the continuous phase
transitions in presence of disorder, the correlation-length critical
exponent $\nu$ verifies $\nu\ge 2/d$, $d$ being the dimensionality of the
space~\cite{Chayes}.

When the pure model shows a first order phase transition the situation
is more complicated. However, there are a set of important results,
both numerical and analytical.  For instance, Aizenman and
Wehr~\cite{Aizenman} showed rigorously that in two dimensions when
introducing disorder, its conjugated density becomes a continuous
function of the thermodynamic parameters. For instance, in a site
diluted model the conjugate density is the energy, while in the Random
Field Ising Model (RFIM), it is the magnetization.  If the phase transition
becomes continuous, one may ask about its Universality Class. A widely
studied model in this context has been the $q$-states Potts model,
whose pure version in $d=2$ undergoes a first order phase transition
for $q\geq 5$.  In recent numerical simulations using Monte Carlo or
transfer matrix methods, the $\nu$ exponent has been found compatible
within errors with the pure Ising value ($\nu=1$) independently of
$q$, but for the magnetic exponent, $\beta$, the numerical results
for $q>2$ are significantly different from the pure Ising
value~\cite{CardyJacobsen2,Chatelain,Chatelain2,Picco2,Young}.

Unfortunately, for $d>2$ the situation is not so clear.
Cardy and Jacobsen~\cite{CardyJacobsen1} (see also Cardy~\cite{Cardy})
have put forward a picture of the general behavior by means of a
mapping from the diluted $q$-states Potts model 
to the RFIM. Their mapping being
asymptotically exact for  large $q$, their results are
also expected to hold for phase transitions with large
latent heat. According to them, when a system that 
undergoes a first-order phase transition gets weakly diluted, the
latent heat decreases. For larger dilutions, it will eventually vanish
at a so-called tricritical point. Cardy and Jacobsen relate the latent
heat with the order parameter of the RFIM, 
whose behavior is governed
by the zero-temperature fixed point. In this way they are able to
predict the critical exponents for the tricritical point
\hbox{$\beta=\beta_{\mathrm{RFIM}}$} and
$\nu=\nu_{\mathrm{RFIM}}/(2-\alpha_{\mathrm{RFIM}} -\beta_{\mathrm
{RFIM}})$\cite{HYPER}. The critical-behavior of the system for larger dilutions
remains unaddressed in their work.  A physical realization of this
scenario  is provided by some magnetic semiconductors like
Zn$_{1-x}$Mn$_{x}$Te~\cite{ZnMnTe}. The magnetic atoms of these
materials behave as  Heisenberg spins living in a fcc lattice, with 
antiferromagnetic interactions. In these highly frustrated systems 
a first order phase transition is found in pure samples that gets
second order upon dilution. 

We finish this overview describing the results obtained by Elderfield
and Sherrington for the diluted Potts Model in the Mean Field
approximation~\cite{Sherrington}. They found that the phase
transition is first order for $q>2$, for all values of
spin-concentration. Another interesting and related model is the Potts
glass for $q>4$ that, according to Mean Field theory, undergoes a first order
transition with no latent heat while for $q=3$ and $4$ the transition is
continuous~\cite{Gross}.

In this work, we shall consider the effects of site-dilution in the
three dimensional three state Potts model, 
whose pure version presents a weak first
order transition (small latent heat or large correlation length at the
critical point).  Our choice has been motivated by the ubiquity of
weak first order transitions in nature.  In particular, the pure $q=3$
Potts model shows very different experimental realizations appearing
in very distant fields. For instance, we can cite the
de-confining phase transition in quenched Quantum
Chromodynamics.  It can also characterize different situations
in solid state physics. For instance a cubic ferromagnet with three
easy axes of magnetization when a magnetic field in the diagonal of
the cubic lattice is turned on (e.g. DyAl$_2$), structural phase
transitions (e.g. SrTiO$_3$) and some fluid mixtures of five (suitable
chosen) components~\cite{Wu}.

We now describe briefly the phase diagram of the three dimensional
three states Potts model in the temperature--concentration
plane $(T,p)$ (see Fig.~\ref{DIAGRAMA}).  The pure model undergoes a (weak)
first order phase transition, at a critical temperature $T_{\mathrm
c}(p=1)$ separating the paramagnetic high-temperature phase from the
low temperature ordered one. This first order transition can be, in
principle, continued inside the $(T,p)$ plane, where the critical
temperature $T_{\mathrm c}(p)$ will lower for smaller $p$. 
The latent-heat for the
first-order phase transition will decrease until the tricritical
point. At this point the model suffers a second order phase transition
that continues (belonging to another Universality Class) until the
$T_{\mathrm c}(p_{\mathrm c})=0$ percolation limit.
We remark that, in the most economic picture, this phase
diagram presents three different Universality Classes: site
percolation in three dimensions (which has been studied in the
literature, e.g. in Ref.~\onlinecite{Perco3D}), the Universality Class of
the tricritical point (conjectured in Ref.~\onlinecite{Cardy}) and the
Universality Class that controls the critical behavior in the region
between the tricritical point and the percolation point.

In this paper we will restrict ourselves to the study of the
second-order line.  Although an experimental realization of the site
diluted Potts model is not yet known (disorder tends to couple with
the order parameter rather than with the energy), whenever it will
appear it will be interesting to have clear theoretical predictions at
hand. The techniques used in this paper are well suited for second
order transitions, but they should be modified in the concentration
range for which the phase transition is first order.  Work is in
progress to study this region~\cite{WorkInProgress}.

\section{The Model and Observables}

We have studied the three dimensional Site Diluted three state Potts
Model, whose Hamiltonian defined on
a cubic lattice with volume $V=L^3$ is
\begin{equation}
{\cal H}= {\mathrm Re} \left[\sum_{<i,j>} \epsilon_i \epsilon_j z_i\,z_j^*
\right]
\, ,
\label{hamiltonian}
\end{equation}
and periodic boundary conditions are applied. In Eq.~(\ref{hamiltonian})
$z_i$'s are complex roots of $ z^3=1$, and 
$\epsilon_i$'s are uncorrelated quenched random variables which are $1$ with
probability $p$, and $0$ with probability $1-p$.  
The Boltzmann weight is proportional to
$\exp(-\beta {\cal H})$.

We have used clusters algorithms in order to update the system.  In a
diluted system, the set of occupied sites can present regions that are
lightly connected to the percolating cluster. These regions are very
difficult to equilibrate just with a single cluster
algorithm~\cite{Wolff}. We have found that a single cluster algorithm
combined with a Heat Bath sweep per measure is efficient for large
concentrations. However, for small concentrations ($p\leq 0.5$) the
previous method is not efficient enough due to the presence of
intermediate-sized clusters, and we have used the Swendsen-Wang
algorithm~\cite{Swendsen}.

We have simulated at $p=1.0,0.9,0.8,0.7,0.6,0.5$ and $0.4$ in lattices
$L=8,16,32$ and $64$. For $p=0.8,0.7,0.4$ we have also run in $L=128$
lattices. We have performed $N_I=200$ nearly independent measures in
every single disorder realization. For $p\leq 0.8$ the number of these
realizations has been $N_S=10000$, except for $p=0.8$, $L=128$, where
we have fixed $N_S=1000$.  In the $p=0.9$ case we have measured in
2000 different disorder realizations.  The total amount of CPU time
has been the equivalent of 16 years of 200 Mhz Pentium-Pro processor.
For small dilutions we have performed the usual $\beta$
extrapolation~\cite{Falcioni} while for $p\leq 0.5$ we used a $p$
extrapolation method~\cite{Perco4D}.  Let us recall that when planning
a disordered model simulation, one should balance two competing
effects. First, to minimize statistical errors, it is better to work
in a $N_I\ll N_S$ regime. On the other hand, if $N_I$ is too small,
the usual calculation of $\beta$ derivatives and extrapolations is
biased. We follow the same procedure of Ref.~\onlinecite{Dilu4D} to
eliminate the bias. With our simulation strategy ($N_I\ll N_S$),
it is crucial to check that the system is sufficiently thermalized
while taking measures. In order to ensure this, we have systematically
compared the results coming from hot and cold starts: half of
our statistics for the largest lattices have been obtained with hot
starts while the other half comes from cold starts.

Regarding the observables, in addition to the energy we have measured
the complex magnetization
\begin{equation}
M=\sum_i \epsilon_i\,z_i\,, 
\end{equation}
from which we obtain the real susceptibility as
\begin{equation}
\chi=\frac{1}{V} 
\overline{\langle |M|^2  \rangle}\ .
\label{chi}
\end{equation}
We have denoted
with $\langle (\cdots) \rangle $ the thermodynamical
average with fixed disorder and with $\overline{ (\cdots)}
$ the average over the disorder.

The formul\ae~for the cumulants read
\begin{eqnarray}
g_2&=&\frac{\overline{\langle |M|^2\rangle^2}-\overline{\langle
|M|^2\rangle}^2}{\overline{\langle |M|^2\rangle}^2}\,,\\
g_3&=&\frac{\overline{\langle M^3\rangle}}{\overline{\langle |M|^2\rangle^{3/2}}}\,,\\
g_4&=& 2 -\frac{\overline{\langle |M|^4\rangle}}
{\overline{\langle |M|^2\rangle^2}}\,,
\label{CUMULANTES}
\end{eqnarray}
$g_4$ being  the standard Binder cumulant, $g_2$ measures whether
the susceptibility is or not a self-averaging quantity and $g_3$ has
been introduced since the three states Potts model is invariant under
a global transformation of the $Z_3$ group. The other cumulants, $g_2$
and $g_4$, are also trivially invariant since we have used the modulus
of the complex magnetization in their construction.

We have used a quotient method~\cite{Quotient}, 
in order to compute the critical exponents. We recall briefly the basis of
this  method. Let $O$ be a quantity diverging 
as $t^{-x_O}$ ($t$ being the reduced temperature) in the
thermodynamical limit. We can write the dependence of $O$ on $L$ and
$t$ in the following way
\begin{equation}
O(L,t)=L^{x_O/\nu} \left[F_O\left(\frac{L}{\xi(\infty,t)}  \right)+
{\cal O}(L^{-\omega},\xi^{-\omega}) \right] ,
\label{obs}
\end{equation}
where $F_O$ is a (smooth) scaling function and $(-\omega)$ is the biggest
non positive eigenvalue of the Renormalization Group
transformation (the corrections-to-scaling exponent). This expression
contains the not directly measurable term $\xi(\infty,t)$, but if
we have a good definition of the correlation length in a finite box
$\xi(L,t)$, Eq.~(\ref{obs}) can be written
\begin{equation}
O(L,t)=L^{x_O/\nu} \left[G_O\left(\frac{\xi(L,t)}{L}  \right)+
{\cal O}(L^{-\omega}) \right] ,
\end{equation}
where $G_O$ is  a smooth function related with $F_O$ and $F_\xi$ and
we have neglected the term $\xi_\infty^{-\omega}$ because we are
working deep in the scaling region.
The definition of the correlation length on a finite box that we use 
is~\cite{Cooper}:

\begin{equation}
\xi=\frac{\sqrt{ \chi/F-1}}{ 2 \sin(\pi/L)} \ ,
\label{correlation}
\end{equation}
where $\chi$ was defined in Eq.~(\ref{chi}) and $F$ is given by
\begin{equation}
F=\frac{V}{3}  \sum_{\Vert \bbox{k}\Vert=\frac{2\pi}{L}} 
\overline{\langle |{\hat z} (\bbox{k}) |^2\rangle }\ ,
\end{equation}
${\hat z}(k_1,k_2,k_3)$ being the discrete Fourier transform of
$\epsilon_i z_i$.
We remark that the definition in Eq.~(\ref{correlation})  makes sense as
a correlation length only in the pure paramagnetic phase of the model.

The main formula of the quotient method is 
\begin{equation}
\left.Q_O\right|_{Q_\xi=s}=\frac{O(sL,t)}{O(L,t)}=
s^{x_O/\nu}+{\cal O}(L^{-\omega})\ ,
\label{QUO}
\end{equation}
e.g. we compute the quotient between $O(sL,t)$ and $O(L,t)$ at the
reduced temperature, $t$, in which $\xi(sL,t)/\xi(L,t)=s$. 
As particular cases
of interest we cite the susceptibility, $\chi$, and the $\beta$-derivative
of the correlation length, $\partial_\beta\xi$,
whose associated exponents are:
\begin{eqnarray}
x_{\partial_\beta\xi}&=&1+\nu\,,\\ 
x_{\chi}&=&(2-\eta)\nu .
\label{OBSERVABLES}
\end{eqnarray}

A clean measure of scale invariance is provided by 
$\left.(\xi/L)\right|_{Q_\xi=s}$.
Let us recall that $\xi/L$ is a monotonically growing function of the
inverse temperature. In the ordered phase it grows as $L^{d/2}$, while
in the disordered phase decreases with growing lattice size. Therefore,
for any pair of lattice sizes, there is a crossing temperature 
where $Q_\xi=2$.  In a second order transition,
$\xi/L$ at the crossing point should tend to a non-vanishing universal
value. For a first order transition, the crossing temperatures tend to
the transition point but $\xi/L$ at the crossing diverges due
to the coexistence of ordered and disordered phases.

We finally analyze the quotient of  the cumulants 
$g_2$, $g_3$ or $g_4$ at two different
lattices, $L$ and $sL$,  
computed at the temperature where $Q_\xi=s$.
Notice  that for a second order phase transition 
the asymptotic limit ($L \to \infty$) of
these quotients is $1$ corrected by terms like
$L^{-\omega}$ (see Eq.~(\ref{QUO})).

The quotient method, Eq.~(\ref{QUO}), has several interesting
features. First, we profit of the large statistical correlation between
$Q_O$ and $Q_\xi$. Next, one does not
need a previous estimate of the infinite volume critical
point. Finally, it allows a simple control of the scaling corrections. 
All of this makes the method specially efficient for the measures of anomalous
dimensions.

\section{Numerical Results}

\begin{figure}
\begin{raggedright}
\leavevmode
\epsfig{file=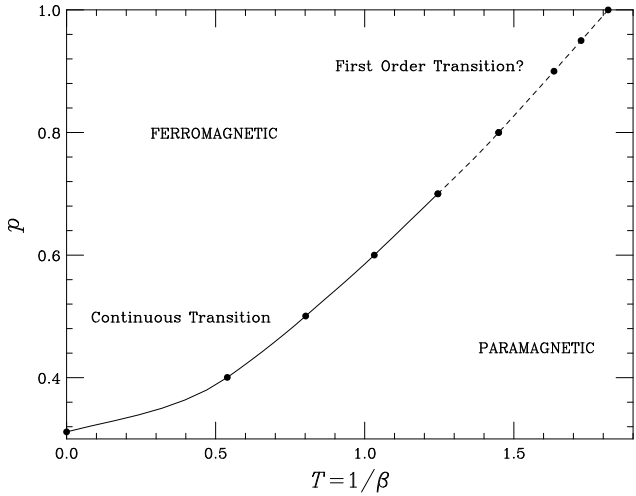,width=0.76\linewidth,angle=90}
\end{raggedright}
\caption{ Phase diagram of the model defined in Eq. (\ref{hamiltonian}). 
The critical line has been drawn using the critical points obtained 
in our numerical simulations (the filled circles in the plot).  
}
\label{DIAGRAMA}
\end{figure}

The phase diagram that we have obtained numerically is shown
in Fig.~\ref{DIAGRAMA} and it agrees with the standard picture.  The
dashed line corresponds to the part of the transition line where we do
not find a clear second order asymptotic behavior. A clear first order
signature is very difficult to see even for dilutions as small as
$p=0.95$. In this region the individual samples usually present a
double peak structure. Consequently the thermalization process is
very slow and a different algorithm for updating must be used.
Work is in progress to analyze this region~\cite{WorkInProgress}.

\begin{figure}
\begin{raggedright}
\leavevmode
\epsfig{file=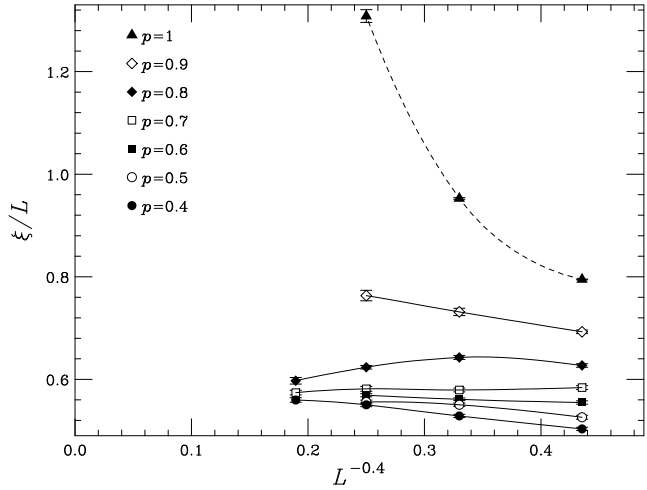,width=0.75\linewidth,angle=90}
\end{raggedright}
\caption{The ratio $\xi/L$ as a function of $L^{-0.4}$ for different
values of the concentration $p$. Notice that the behavior for $p\geq
0.9$ is clearly different from the rest of concentrations, where a
non-vanishing, dilution-independent limiting value is likely to
occur.}
\label{XIL}
\end{figure}

Our scope is now to compute the critical exponents in the region in
which the transition is clearly second order, i.e. the study of the
Universality Class between the tricritical and percolation limits.
The first stage is to determine where an asymptotic second order
behavior has been reached with lattice sizes up to $L=128$.  In
Fig.~\ref{XIL} we show the value of $\xi/L$, at the points for 
which $Q_\xi=2$ for
the different $(L,2L)$ lattice pairs and as functions of
$L^{-\omega}$.

We have used for $\omega$ the corresponding value of
the site diluted Ising model~\cite{Dilu3D}.  For $p\leq 0.7$ we find
that $\xi/L$ seems to tend to a dilution-independent
value. Notice the clear divergence for $p=1$, where the transition
is known to be first order.  For $p=0.9$ we find a similar trend that
for the pure case while for $p=0.8$ we find a transient behavior: for
small lattices $\xi/L$ grows while in the largest lattices it seems to
approach the universal value.

\begin{figure}
\begin{raggedleft}
\epsfig{file=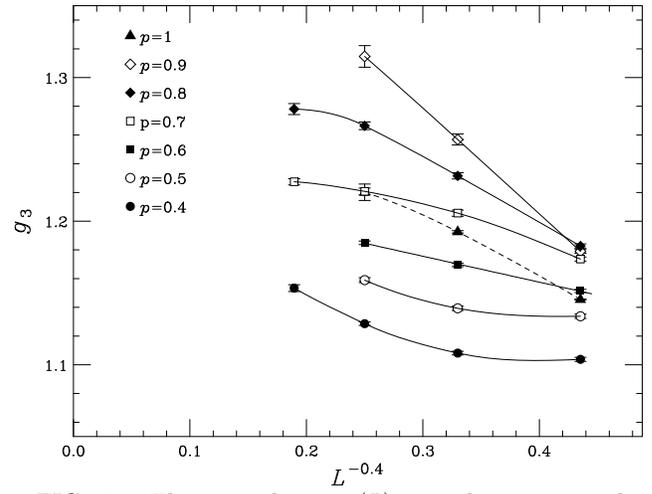,width=0.75\linewidth,angle=90}
\end{raggedleft}
\caption{The cumulant $g_3(L)$ at the points where
$\xi(2L,t)=2\xi(L,t)$ as a function of $L^{-0.4}$ for the 
different concentrations considered.}
\label{G3}
\end{figure}

Another interesting quantity in order to clarify the second order
behavior is the cumulant $g_3$. The $g_3(L)$ values at the points
where $\xi(2L,t)=2\xi(L,t)$ are shown in Fig.~\ref{G3}.  In this case
we see a different scaling behavior for $p=0.9$ and $p=0.8$
up to the studied lattice sizes.  We also guess from this figure 
that the $\omega$ value cannot be much larger than 0.4.

\begin{figure}
\begin{raggedleft}
\leavevmode
\epsfig{file=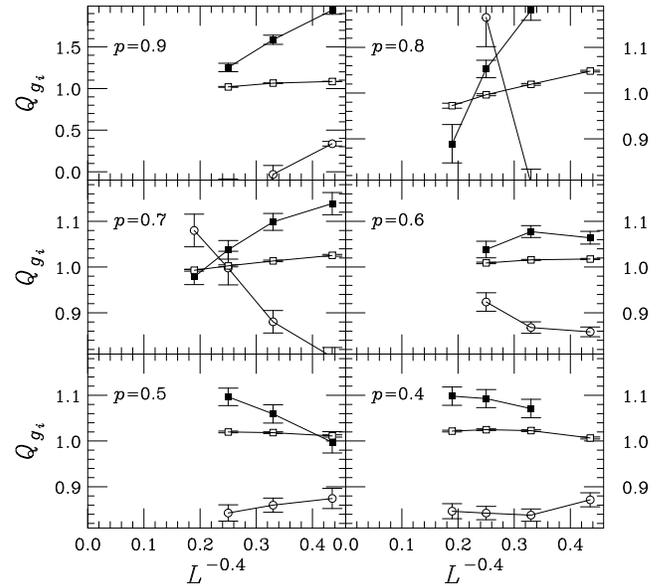,width=0.9\linewidth,angle=90}
\end{raggedleft}
\caption{Quotients of the cumulants $g_2$, $g_3$ and $g_4$ (filled squares, 
open squares and open circles, respectively) as a function of
$L^{-0.4}$. Notice the different $y$ scale in the $p=0.9$ case.}
\label{Qptot}
\end{figure}

We have next considered the quotients of the different cumulants $g_i$
at the points where $Q_\xi=2$. We recall that these quantities should
go to 1 as $L$ tends to infinity in a second order phase transition.
We present our results in Fig.~\ref{Qptot}.  At concentrations $p=0.9$
and $p=0.8$ we do not find an asymptotic behavior. For $p=0.7$, the
behavior is yet not monotonous. Only for $p=0.4,0.5$ and $0.6$ it
seems that the asymptotic behavior is reached. Unfortunately, a
reliable estimate of $\omega$ cannot be obtained but our results point
to a value near $0.4$. Moreover, the higher
order scaling corrections are rather strong for these quantities. Finally,
let us remark that the corrections to scaling and statistical errors
are much larger for $g_2$ and $g_4$ than for $g_3$.

Therefore, for the study the second order region, we conclude that only
for $p\leq 0.6$ an asymptotic scaling behavior for the
considered lattice sizes has been found.

\vbox{
\narrowtext
\begin{table}[h]
\begin{tabular}{rllllll}
$L$ & $p=0.9$ & {$p=0.8$} & {$p=0.7$} & {$p=0.6$} & {$p=0.5$} &
{$p=0.4$} \\\hline
8  &0.571(4) &  0.633(3) &  0.662(4)& 0.685(3)& 0.706(4) & 0.738(5) \\
16 &0.592(7) &  0.659(3) &  0.686(3)& 0.692(3)& 0.698(4) & 0.711(5) \\
32 &0.664(12)&  0.700(4) &  0.695(4)& 0.688(4)& 0.694(4) & 0.696(4) \\
64 &   ---   &  0.711(13)&  0.707(4)&  ---   &   ---    & 0.692(4) \\
\end{tabular}
\caption{Apparent critical exponent $\nu$, obtained from
$Q_{\partial_\beta \xi}$ measured where $Q_\xi=2$ for all the concentrations
studied.}
\protect\label{table:nu}
\end{table}
}

\vbox{
\narrowtext
\begin{table}[h]
\begin{tabular}{rcccccc}
$L$ & $p=0.9$ & $p=0.8$ & $p=0.7$ & $p=0.6$ & $p=0.5$ & $p=0.4$ \\\hline
8 &  0.048(2)&   0.057(2)&  0.065(2)& 0.0745(15)& 0.079(4)&  0.072(4)\\
16&  0.036(4)&   0.045(2)&  0.068(2)& 0.0773(14)& 0.079(2)&  0.077(3)\\
32&--0.029(9)&   0.050(3)&  0.074(3)& 0.077(2)  & 0.079(2)&  0.077(3)\\
64&    ---   &   0.064(6)&  0.071(3)&  ---      &  ---    &  0.080(3) \\
\end{tabular}
\caption{Apparent critical exponent  $\eta$, computed using
$Q_\chi$, for all the considered concentrations.}
\protect\label{table:eta}
\end{table}
}

We report the results for the critical exponents as functions of $p$
and $L$ in tables~\ref{table:nu} and~\ref{table:eta}.  We have applied
Eq.~(\ref{QUO}) with $s=2$ to $\partial_\beta \xi$ for computing $\nu$
and to $\chi$ for extracting $\eta$.  We can observe that the asymptotic
behavior of these estimates for $p \ge 0.7$ is not clear. 

\begin{figure}
\begin{raggedleft}
\leavevmode
\epsfig{file=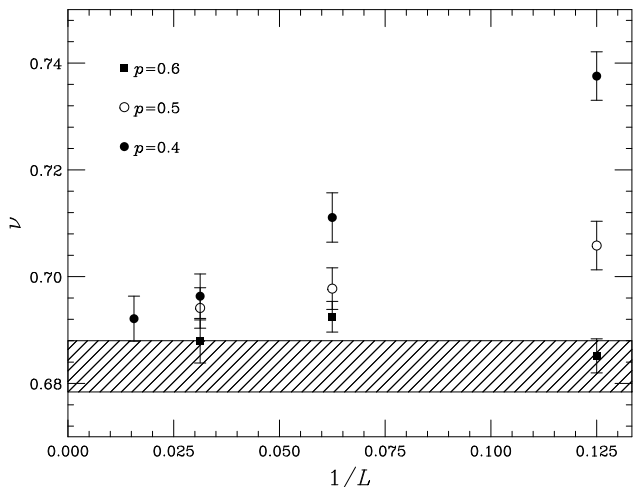,width=0.75\linewidth,angle=90}
\end{raggedleft}
\caption{ The $\nu$ apparent critical exponent for 
three different concentrations ($p=0.6, 0.5$ and $0.4$) as a function 
of $1/L$.  The shadowed region corresponds to the value of the three 
dimensional site diluted Ising model~\cite{Dilu3D}.}
\label{NU}
\end{figure}

\begin{figure}
\begin{center}
\leavevmode
\epsfig{file=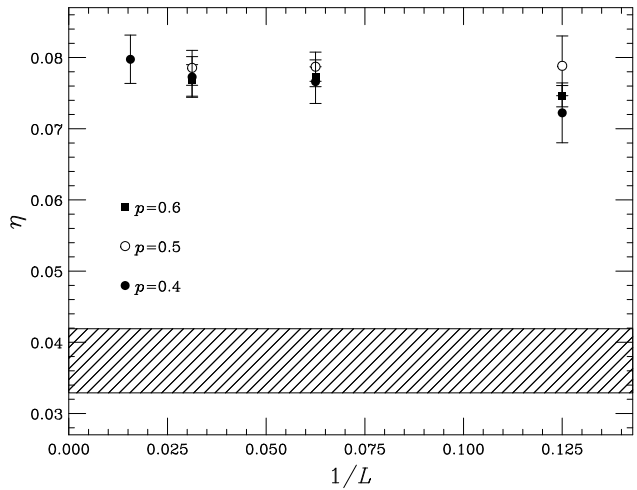,width=0.75\linewidth,angle=90}
\end{center}
\caption{The finite size estimate for $\eta$ critical exponent for 
three different concentrations ($p=0.6, 0.5$ and $0.4$) as a function 
of $1/L$.  We show the value corresponding to the three dimensional
site diluted Ising model~\cite{Dilu3D} with a shadowed region.}
\label{ETA}
\end{figure}

We have plotted the $\nu$ and $\eta$ apparent critical exponents for
$p=0.6,0.5$ and $0.4$ as functions of $1/L$ in Figs.~\ref{NU}
and~\ref{ETA} respectively. We recall that we have found a
$\omega\approx 0.4$ value for the cumulants. From the Figs.~\ref{NU}
and~\ref{ETA} we see that $\omega=1$ for the leading
scaling-corrections term could be a reasonable choice in this case.  A
possible explanation of this contradiction could be that for the
observables used for computing the critical exponents the leading term
($\omega \approx 0.4$) vanishes. In any case we should remark that we
have not a precise control over the scaling corrections unlike, for
example, in the investigation of the three dimensional site diluted
Ising model~\cite{Dilu3D}. Fortunately, the scaling corrections for
the critical exponents are rather small. Thus, it is not essential in
this model to perform an infinite-volume extrapolation of our
estimates. This is in marked contrast with the Ising case, where the
extrapolation procedure was crucial to correctly compute the critical
exponents.

From Figs.~\ref{NU} and \ref{ETA} we estimate the critical exponents
in the second order region as the displayed in the last row of the 
table~\ref{table:models}. We remark that the $\nu$ value is indistinguishable
from the 3D site diluted Ising model one. However, the $\eta$ value is very 
different from the values found for the rest of the models reported in
the table~\ref{table:models}.

\vbox{
\narrowtext
\begin{table}[h]
\begin{tabular}{llll}
Model             &     $\nu$          & $\eta$    & $\beta$  \\ \hline
Pure Ising~\cite{Perco3D}              & 0.6294(10)&0.0374(12)&0.3265(4)\\
Diluted Ising~\cite{Dilu3D}            & 0.684(5)  &0.037(5)  &0.355(3)\\
Percolation~\cite{Perco3D}             & 0.8765(18)&--0.0460(3)&0.4181(9)\\ 
Tricritical Point~\cite{TRIC}          & 0.68(5)  &0.50(5)    &0.00(5)  \\ 
This work              &0.690(5)& 0.078(4)
& 0.372(4)\\
\end{tabular}
\protect\label{table:models}
\caption{Critical exponents for some three dimensional models.}
\end{table}
}

\section{Conclusions}

We have numerically studied the three dimensional site diluted three
states Potts model. The phase diagram in the temperature-concentration
plane consists of a ferro-magnetically ordered phase separated from
a paramagnetic, high temperature one. Between both regions there is a 
critical line, which is (weakly) first order in the limit of pure
samples. For small concentrations, a clear second-order behavior
is found, while the region with $p\gtrsim 0.9$ shows a different
behavior, probably corresponding to a cross-over, more difficult to
analyze.
  
The critical exponents of the second order region have been computed
using Finite Size Scaling Techniques.  We have found that these
exponents are dilution independent, and that they show a very mild
evolution with the lattice size. That is why a sound estimate of the
critical exponents can be given, in spite of the fact that we have
been unable of measuring the scaling-corrections exponent $\omega$.
This is in marked contrast with the situation in the site-diluted
Ising model, where the scaling-corrections are severe but $\omega$ can
be obtained with a $15\%$ accuracy.

Regarding the variation of the critical exponents with $q$ we have
compared the results for the Potts model and the Ising case.  We have
found that the $\nu$ exponent varies slowly (or perhaps remains
unchanged) with the $q$ value whereas the $\beta$ or $\eta$ exponents
show a clear variation on this parameter.  This picture strongly
reminds the obtained in numerical simulations in two dimensions.
 
The study of the first-order region and of the critical behavior
in the neighborhood of the tricritical point requires rather different
numerical techniques and will be the matter of future work.

\section{Acknowledgments}

We gratefully acknowledge discussions with D. Belanger, J. Cardy and
H. Rieger.  We thank partial financial support from CICyT (AEN97-1708
and AEN97-1693).  The computations have been carried out using the
RTNN machines (Universidad de Zaragoza and Universidad Complutense de
Madrid) and the ORIGIN2000 at the Centro de Supercomputaci\'on
Complutense (CSC).

\end{multicols}

\begin{thebibliography}{99}
\bibitem{HYPER} 
This formula is obtained from the equation~\cite{CardyJacobsen1}
\begin{equation}
1/\nu=d-\theta_{\mathrm
RFIM} -\beta_{\mathrm RFIM}/\nu_{\mathrm RFIM}\label{CJ}
\end{equation}
using the modified hyperscaling relation 

\begin{equation}
\alpha_{\mathrm RFIM}=2-(d-\theta_{\mathrm RFIM}) \nu_{\mathrm RFIM}\,,
\label{MODHYPER}
\end{equation}
where $d$ is the
spatial dimension and $\theta_{\mathrm RFIM}$ the exponent which controls
the violation of hyperscaling.
\bibitem{TRIC}
The experimental data\cite{BELANGER_BOOK} ($\alpha_{\mathrm RFIM}\simeq 0$, 
$\nu_{\mathrm RFIM}\simeq 1$, $\gamma_{\mathrm RFIM}\simeq 1.7$ and 
${\bar \gamma}\simeq 3$)
seem to support the modified hyperscaling relation (\ref{MODHYPER})
but the numerical simulations do not, neither
with binary random fields nor with Gaussian ones. To
compute the
critical exponents of the tricritical point we use Eq.~(\ref{CJ}) 
with the values of $\theta_{\mathrm RFIM}=1.53(1)$,
$\beta_{\mathrm RFIM}=0.00(5)$ and $\nu_{\mathrm RFIM}=1.1(2)$
obtained in numerical
simulations with Gaussian random fields\cite{Rieger}, 
but we  ignore the $\alpha_{\mathrm
RFIM}$ value, what allows to avoid the use of the
modified hyperscaling relation.
\end{thebibliography}
\end{document}